\documentclass[fleqn,usenatbib]{mnras}

\usepackage[T1]{fontenc}

\DeclareRobustCommand{\VAN}[3]{#2}
\let\VANthebibliography\thebibliography
\def\thebibliography{\DeclareRobustCommand{\VAN}[3]{##3}\VANthebibliography}

\usepackage{graphicx}	%
\usepackage{amsmath}	%
\usepackage{amssymb}	%
\usepackage[flushleft]{threeparttable}
\usepackage{newtxtext,newtxmath}
\usepackage{float}
\newcommand{\kms}{\,km\,s$^{-1}$}%

\title[Finding magnetic north]{Finding magnetic north: an extraordinary magnetic field detection in Polaris and first results of a magnetic survey of classical Cepheids}
\author[J. A. Barron et al.]{
J. A. Barron,$^{1,2}$\thanks{Based on observations obtained at the Canada-France-Hawaii Telescope (CFHT) which is operated by the National Research Council (NRC) of Canada, the Institut National des Sciences de l'Univers of the Centre National de la Recherche Scientifique (CNRS) of France, and the University of Hawaii. The observations at the CFHT were performed with care and respect from the summit of Maunakea which is a significant cultural and historic site.}\thanks{E-mail: j.barron@queensu.ca}
G. A. Wade$^{2}$,
N. R. Evans$^{3}$,
C. P. Folsom$^{4}$
and H. R. Neilson$^{5}$
\\
$^{1}$Department of Physics, Engineering Physics \& Astronomy, Queen’s University, 64 Bader Lane, Kingston, ON K7L 3N6, Canada\\
$^{2}$Department of Physics and Space Science, Royal Military College of Canada, PO Box 17000, Kingston, ON K7K 7B4, Canada\\
$^{3}$Smithsonian Astrophysical Observatory, MS 4, 60 Garden St., Cambridge, MA 02138, USA\\
$^{4}$Tartu Observatory, University of Tartu, Observatooriumi 1, Tõravere, 61602 Tartumaa, Estonia\\
$^{5}$David A. Dunlap Department of Astronomy \& Astrophysics, University of Toronto, 50 St. George Street, Toronto, ON M5S 3H4, Canada\\
}

\date{Accepted XXX. Received YYY; in original form ZZZ}

\pubyear{2021}

\begin{document}
\label{firstpage}
\pagerange{\pageref{firstpage}--\pageref{lastpage}}
\maketitle

\begin{abstract}
Classical Cepheids are essential objects in the study of stellar evolution and cosmology; however, we know little about their magnetic properties. We report the detection of Stokes $V$ features interpreted as Zeeman signatures in four classical Cepheids using high-resolution spectropolarimetric observations obtained with ESPaDOnS at CFHT. Eight observations of $\eta$~Aql were acquired in 2017 covering its 7.2\,d pulsation period, and single observations of Polaris, $\zeta$~Gem, $\delta$~Cep and RT~Aur were obtained in 2020 as part of our ongoing systematic survey. We use mean circular polarization Stokes~$V$ profiles generated using the Least-Squares Deconvolution procedure to diagnose Zeeman signatures and measure mean longitudinal field strengths $\langle B_{z}\rangle$. We detect magnetic signatures across all pulsation phases of $\eta$~Aql ($-0.89\pm0.47$\,G$\,<\langle B_{z}\rangle<1.27\pm 0.40$\,G), as well as in the single observations of Polaris ($0.59\pm0.16$\,G), $\zeta$~Gem ($0.41\pm0.16$\,G) and $\delta$~Cep ($0.43\pm0.19$\,G). The Stokes~$V$ profile of Polaris is detected at extremely high S/N and implies a complex magnetic field topology. It stands in stark contrast to all other detected Stokes~$V$ profiles, which show unusual approximately unipolar positive circular polarization lobes analogous to those observed in some Am stars.
\end{abstract}

\section{Introduction}
Classical Cepheid variables ("Cepheids") are powerful astrophysical tools in the study of both stellar evolution and cosmology due to their regular pulsation and tightly correlated period-luminosity relation (Leavitt law;  \citealt{leavitt_1912}). Cepheids are luminous yellow supergiants, the helium core-burning descendants of main-sequence B-type stars, and populate the classical instability strip of the Hertzsprung-Russell diagram. Variations in temperature and luminosity during their radial pulsations give rise to their characteristic light curves. During pulsation the extended atmospheres of Cepheids undergo complex dynamics and exhibit poorly understood phenomena including shocks \citep{hocde_2020}, pulsation modulated UV and X-ray emission \citep{engle_2014, engle_2017}, and the production of circumstellar envelopes \citep{nardetto_2016}.

Surprisingly little is known about the magnetic properties of Cepheids and what role magnetic fields may play in their atmospheric structure and evolutionary paths. It has been proposed that magnetic fields may modulate the pulsation of short-period Cepheids (e.g. Polaris) by inhibiting convection \citep{stothers_2009}. Phase-modulated UV and X-ray emission has been detected in some Cepheids; this has been proposed to be due to shock waves or magnetic reconnection events during pulsation \citep{engle_2017, moschou_2020}. Historically, there have been several studies aimed at the direct detection of Cepheid magnetic fields (e.g. \citealt{borra_1981, borra_1984, plachinda_2000, wade_2002}), however, there exist few reliable magnetic field measurements from modern high-resolution spectropolarimetry. The first convincing Cepheid magnetic detection comes from a spectropolarimetric survey of massive, late-type supergiants by \cite{grunhut_2010} using ESPaDOnS at the Canada-France-Hawaii Telescope (CFHT). The authors detected a Zeeman signature in $\eta$~Aql with a weak $\sim\!1$\,G longitudinal field. Similar $\sim\!1$\,G fields were found in about a third of the supergiant sample, and no magnetic field was detected in the bright Cepheids $\zeta$~Gem and $\delta$~Cep. These results indicate that current generation spectropolarimeters (e.g. ESPaDOnS) are necessary to obtain high precision magnetic measurements of Cepheids.

Motivated by the lack of direct observational constraints on the characteristics of Cepheid magnetic fields, we aim to conduct a first systematic spectropolarimetric survey of the brightest $\sim\!20$ Galactic Cepheids. This survey will provide a first characterization of Cepheid magnetic properties by obtaining a single observation of each target at the highest feasible magnetic precision. We seek to answer questions regarding the detectability, incidence and topology of Cepheid magnetic fields. In addition, we aim to gain first insights into the relationship between Cepheid magnetic fields and atmospheric dynamics (e.g. X-ray emission, stellar winds) and the role of magnetic fields in Cepheid evolution (e.g. magnetic braking, enhancement or confinement of mass loss, suppression of convection). Here we describe results obtained for the first five targets observed in our survey.

\section{Observations}
All observations\footnote{CFHT program IDs 17BC19 (PI: Wade) and 20BC20 (PI: Barron)} were obtained using ESPaDOnS, a fibre-fed echelle spectropolarimeter mounted at CFHT. It has a high resolving power of $R\!\sim\!65,000$ and spans a spectral range of $370\!-\!1050$\,nm over 40 overlapping spectral orders. Each complete observation consists of four polarimetric sub-exposures, between which the instrument's Fresnel rhombs are rotated. This yields an unpolarized intensity (Stokes~$I$) spectrum, a circularly polarized (Stokes~$V$) spectrum, and two diagnostic null ($N$) spectra which are used to identify spurious signals in Stokes~$V$. Polarized Stokes~$V$ spectra are sensitive to the Zeeman effect \citep{donati_2009}, which allows us to detect and characterize stellar photospheric magnetic fields.

Observations of $\eta$ Aql were obtained over eight consecutive nights in 2017 covering its full 7.2 d pulsation cycle. The other four targets (Polaris, $\delta$~Cep, $\zeta$~Gem and RT~Aur) were each observed once in 2020. As the brightness of a Cepheid can vary by over a magnitude during pulsation, we obtained our observations close to maximum brightness ($\phi_{\textnormal{puls}}\!=\!0$) to increase the probability of detection. To reach a suitable signal-to-noise ratio (S/N) multiple spectra were acquired sequentially and co-added using dedicated IDL tools. The spectra were finally interactively rectified using polynomial fits to continuum of each order. See \cite{wade_2016} for a full description of the reduction and analysis of ESPaDOnS data. 

We also analyzed archival ESPaDOnS and NARVAL circular polarization observations of our targets that we retrieved using the PolarBase database \citep{petit_2014}. NARVAL is a twin instrument of ESPaDOnS operating at the T\'elescope Bernard Lyot at Pic du Midi. This archival data includes three observations of $\eta$~Aql (including the detection of \citealt{grunhut_2010}), one observation of $\zeta$~Gem and three observations of $\delta$~Cep. These observations were processed and analyzed in the same manner as the new data.

A log of all Stokes $V$ observations analyzed in this paper is provided in Table~\ref{tab:obs}.

In addition to the Stokes $V$ spectra, we also obtained linear polarization Stokes $Q$ and $U$ spectra using ESPaDOnS for each of our detected targets. These data - obtained to better understand potential origins of the Stokes $V$ features observed in LSD profiles - are described in Appendix~\ref{sect_linear}.  

\begin{table*}
\caption{Log of spectropolarimetric observations and magnetic results. Observations were obtained by ESPaDOnS at CFHT unless otherwise indicated. Included are each target's ID, calendar observation date, heliocentric Julian date at midpoint of observation, pulsation phase $\phi_{\textnormal{puls}}$ at midpoint of observation (adopted ephemerides described in text), exposure time of polarimetric sequence, S/N per 1.8\kms pixel at 500\,nm of co-added spectrum, adopted line mask ($T_{\textnormal{eff}}$, $\log g$), binned velocity width, adopted integration limits, mean longitudinal magnetic field strength $\langle B_{z}\rangle$, false alarm probability (FAP) and detection flag of LSD Stokes~$V$ profile (definitely detected (DD), marginally detected (MD) or not detected (ND)).}
\label{tab:obs}
\begin{tabular}{cccccccccccc}
\hline
Star&Date&HJD\,$-$&$\phi_{\textnormal{puls}}$&Exp. Time&S/N&Mask&dv&Int. lims&  $\langle B_{z} \rangle\pm\sigma_{B}$ &FAP&Det.\\
& &2,400,000&&(s)&&&(km\,s$^{-1}$)&(km\,s$^{-1}$)&(G)&&Flag\\
\hline
\multicolumn{12}{c}{\underline{ New Observations }}\\
Polaris & 2020-11-21 &59174.78629&0.00&$57\!\times\!4\!\times\!11$&6820&T6000g2.0&1.8&$-48.0, 13.2$&$0.59\pm0.16$&$<\!10^{-16}$&DD\\
$\eta$~Aql&2017-08-08&57973.82954&0.99&$16\!\times\!4\!\times\!50$&3180&T5750g2.0&3.6&$-69.6, 9.6$&$0.71\pm0.59$&$8.9\!\times\!10^{-7}$&DD\\
&2017-08-09&57974.80505&0.12&$16\!\times\!4\times\!50$&3630&T5750g2.0&3.6&$-80.4, 9.6$&$1.16\pm0.57$&$1.1\!\times\!10^{-16}$&DD\\
&2017-08-10&57975.81438&0.26&$16\!\times\!4\!\times\!50$&4660&T5750g2.0&3.6&$-62.4, 13.2$&$0.74\pm0.32$&$<\!10^{-16}$&DD\\
&2017-08-11&57976.96474&0.42&$14\!\times\!4\!\times\!50$&3840&T5750g2.0&3.6&$-62.4, 16.8$&$1.45\pm0.36$&$<\!10^{-16}$&DD\\
&2017-08-12&57977.75812&0.53&$16\!\times\!4\!\times\!50$&3980&T5750g2.0&3.6&$-44.4, 24.0$&$0.30\pm0.27$&$<\!10^{-16}$&DD\\
&2017-08-13&57978.80601&0.68&$16\!\times\!4\!\times\!50$&3430&T5750g2.0&3.6&$-40.8, 49.2$&$0.63\pm0.43$&$<\!10^{-16}$&DD\\
&2017-08-14&57979.78417&0.81&$13\!\times\!4\!\times\!50$&3370&T5750g2.0&3.6&$-40.8, 49.2$&$-0.89\pm0.47$&$3.1\!\times\!10^{-6}$&DD\\
&2017-08-15&57980.80505&0.96&$16\!\times\!4\!\times\!50$&5100&T5750g2.0&3.6&$-76.8, 9.6$&$1.27\pm0.40$&$<\!10^{-16}$&DD\\
$\zeta$~Gem&2020-11-30&59184.07391&0.07&$45\!\times\!4\!\times\!43$&4700&T5750g1.5&1.8&$-37.2, 20.4$&$0.41\pm0.16$&$7.8\!\times\!10^{-11}$&DD\\
$\delta$~Cep&2020-12-04&59187.74261&0.94&$48\!\times\!4\!\times\!36$&6530&T6500g2.0&3.6&$-62.4, -1.2$&$0.43\pm0.19$&$1.2\!\times\!10^{-7}$&DD\\
RT~Aur & 2020-12-07 &59191.06000&0.09&$18\!\times\!4\!\times\!190$&4190&T6500g2.0&1.8&$-24.6, 36.6$&$0.03\pm0.24$&$7.3\!\times\!10^{-1}$&ND\\\\
\multicolumn{12}{c}{\underline{ Archival Observations }}\\
$\eta$ Aql&2009-09-05$^{\textrm{a}}$&55079.77405&0.74&$3\!\times\!4\!\times\!111$&1990&T5750g2.0&3.6&$-37.2, 42.0$&$0.57\pm0.59$&$3.6\!\times\!10^{-8}$&DD\\
&2015-07-08$^{\textrm{b}}$&57211.60973&0.78&$3\!\times\!4\!\times\!305$&710&T5750g2.0&3.6&$-30.0, 42.0$&$-0.07\pm0.54$&$4.4\!\times\!10^{-1}$&ND\\
$\zeta$ Gem&2009-09-10$^{\textrm{a}}$&55085.04496&0.22&$3\!\times\!4\!\times\!100$&1750&T5750g1.5&1.8&$-15.6, 24.0 $&$-0.03\pm0.25$&$4.9\!\times\!10^{-1}$&ND\\
$\delta$ Cep&2009-10-02$^{\textrm{a}}$&55106.92239&0.48&$3\!\times\!4\!\times\!97$&1280&T6500g2.0&3.6&$-40.8, 9.6$&$-0.18\pm0.59$&$9.0\!\times\!10^{-1}$&ND\\
&2011-07-15&55758.04518&0.82&$11\!\times\!4\!\times\!21$&1930&T6500g2.0&3.6&$-30.0, 34.8$&$-0.04\pm0.50$&$5.6\!\times\!10^{-1}$&ND\\
&2016-09-01$^{\textrm{b}}$&57632.61567&0.14&$1\!\times\!4\!\times\!150$&380&T6500g2.0&3.6&$-55.2, -8.4$&$-0.50\pm0.93$&$6.9\!\times\!10^{-1}$&ND\\
\hline
\multicolumn{12}{l}{$^\textrm{a}$ Observation published in \cite{grunhut_2010}.}\\
\multicolumn{12}{l}{$^\textrm{b}$ Observation obtained by NARVAL at TBL.}
\end{tabular}
\end{table*}

\section{Magnetic Field Diagnosis}\label{sec:magnetic_diagnosis}
We performed a magnetometric analysis on the Stokes~$V$ spectra following the general procedure of \cite{grunhut_2017}. To detect weak Zeeman signatures we applied the Least Squares Deconvolution (LSD; \citealt{donati_1997}) multi-line analysis procedure to each co-added spectrum using the iLSD routine \citep{kochukhov_2010}. This analysis gives a high S/N mean circular polarization profile (LSD Stokes~$V$), a mean unpolarized profile (LSD Stokes~$I$) and a mean diagnostic null profile (LSD~$N$) for each observation. The line masks used in the LSD computation were generated with an `extract stellar' request from the Vienna Atomic Line Database (VALD3; \citealt{piskunov_1995, ryabchikova_2015}). We assumed solar abundances and literature stellar parameters for each target and excluded all lines with predicted normalized depths less than 10\% of continuum. 

Multiphase parameter studies have shown that Cepheid effective temperature $T_{\textnormal{eff}}$ correlates strongly with phase, and that both surface gravity $\log g$ and microturbulence $V_{t}$ can have slight phase dependencies (e.g. \citealt{luck_2018}). In particular, $T_{\textnormal{eff}}$ can vary by over 1000\,K during a pulsation cycle, and the variations increase in amplitude with increasing period. We adopted parameter values corresponding to the phase of observation for each single observation of Polaris, $\zeta$~Gem, $\delta$~Cep and RT~Aur where $\phi_{\textnormal{puls}}\approx0$ and $T_{\textnormal{eff}}$ is near maximum. We assumed mean stellar parameters for all observations of $\eta$ Aql to ensure consistency of analysis across the observed phases.

 Each generated line mask was then iteratively `cleaned' and `tweaked' to remove contaminating hydrogen and telluric lines using an interactive graphical tool \citep{grunhut_2017}, leaving approximately $4000-6000$ lines per mask. The final LSD profiles were computed with velocity bins of 1.8\kms or 3.6\kms depending on the noise of the Stokes~$V$ profile.
 
 We use the detection of signal in the LSD Stokes~$V$ profile as the key indicator of a photospheric magnetic field. The false alarm probability (FAP; \citealt{donati_1992}) of each Stokes~$V$ profile was calculated to quantify the likelihood of detection. A Stokes $V$ signature is definitely detected (DD) for $\textnormal{FAP}\!<\!10^{-5}$, marginally detected (MD) for $10^{-5}\!<\!\textnormal{FAP}\!<10^{-3}$, and not detected (ND) for $\textnormal{FAP}\!>\!10^{-3}$ \citep{donati_1997}. Several of the $N$ profiles show a visible signature (e.g. $\delta$~Cep; Fig.~\ref{fig:3_lsdplot}), however all are formally ND. This type of null signature has been reported in magnetic $\beta$~Cep stars (e.g. \citealt{neiner_2012, shultz_2017}) and is explained as the result of radial velocity (RV) variations during or between sub-exposures.
 
 The mean longitudinal magnetic field $\langle B_{z}\rangle$ (averaged over the visible portion of the star) was measured from each LSD Stokes~$V$ profile using the first-order moment method \citep{donati_1997, wade_2000}. Note that a significant signal can be detected in a Stokes~$V$ profile even when $\langle B_{z}\rangle$ is not significantly detected. This is because $\langle B_{z}\rangle$ is a disc-averaged measurement and may be equal to zero due to a combination of the magnetic field configuration, rotational phase and rotation axis inclination.
 
 The first-order moment method of measuring the longitudinal magnetic field \citep{1979A&A....74....1R} assumes a null net flux integrated across the Stokes $V$ profile. We show in Sect.~\ref{sec:results} that this is not the case for $\eta$~Aql, $\zeta$~Gem and $\delta$~Cep. It is unclear if the inferred value of $\langle B_{\rm z}\rangle$ is meaningful. Nevertheless, given the importance of the longitudinal field uncertainty as a metric of magnetic sensitivity, we report $\langle B_{\rm z}\rangle$ values for all spectra in Table~\ref{tab:obs}.
 
Bounds for the FAP analysis and $\langle B_{z}\rangle$ measurement are typically set at the continuum limits of the Stokes~$I$ profile, which are expected to coincide with the continuum limits of the Stokes~$V$ profile. However, a number of our Stokes~$V$ profiles have excess flux outside the bounds of the Stokes~$I$ profiles. While the origin of this excess flux is unclear, when present we choose to include it in our analysis. We set each profile's bounds visually, choosing an asymmetric window about the Stokes~$I$ line minimum that incorporates the total Stokes~$V$ flux. The same bounds are used in both the FAP and $\langle B_{z}\rangle$ analysis for each observation and are recorded in Table \ref{tab:obs}. 

\section{Results}\label{sec:results}

\subsection{Polaris}
\begin{figure}
	\includegraphics[width=\columnwidth]{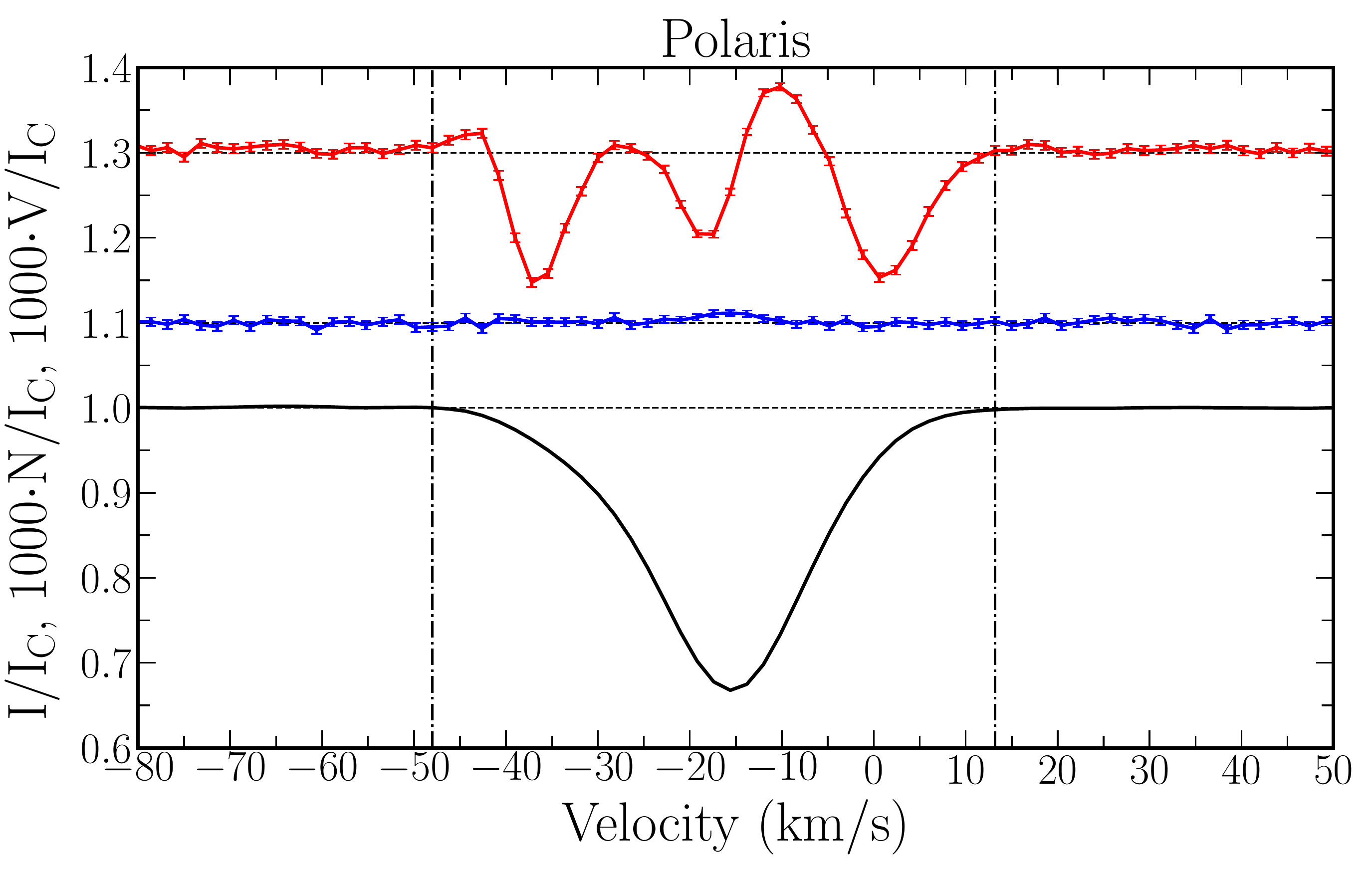}
    \caption{LSD Stokes~$V$ (top, red), diagnostic null $N$ (middle, blue) and Stokes~$I$ (bottom, black) profiles for Polaris. The Stokes~$V$ and diagnostic null profiles have been scaled and offset for visibility. Vertical dash-dotted lines show the velocity bounds used in the magnetic analysis.}
    \label{fig:polaris_lsd}
\end{figure}
Polaris ($\alpha$ UMi, $P_{\textnormal{puls}}=3.97$\,d, $\langle V\rangle=1.98$\footnote{Mean visual magnitudes from \cite{fernie_1995}.}, F7Ib-F8Ib\footnote{Spectral types from the GCVS \citep{samus_2017}.}), the North Star, is an enigmatic object. It has been the subject of observation for centuries due to its role in celestial navigation and is arguably the most well-known star in the night sky. As the nearest and brightest Cepheid, Polaris is an important stellar laboratory in the study of Cepheid structure and evolution. Yet, its mass and evolutionary state remain debated due to disagreements about its distance and a discrepancy between its dynamically measured mass and that inferred from stellar evolution modelling (\citealt{evans_2018} and references therein). Polaris Aa the Cepheid is a member of a triple system, composed of the close astrometric binary Polaris Aa-Ab and its bright visual companion Polaris B ($V=8.2$; F3V) separated by about 18" \citep{evans_2018}. As a short period Cepheid, Polaris exhibits sinusoidal, low amplitude photometric and RV variations \citep{anderson_2019}. Most puzzling, Polaris has a long history of period changes and amplitude variability that are not easily explained (\citealt{anderson_2019} and references therein).

Despite Polaris being both the closest and brightest Cepheid, there have been few investigations into its magnetic properties. \cite{borra_1981} performed seven measurements of Polaris' longitudinal field, finding unsigned field strengths of order $1\!-\!10$\,G and only detection above $3\sigma$. \cite{usenko_2010} provides an additional 25 longitudinal field measurements of Polaris between 1983 and 2005 performed using several different techniques. The unsigned longitudinal field strengths were reported to vary by up to $\sim\!100$\,G, an order of magnitude higher than that found by \cite{borra_1981}. It is difficult to interpret these measurements due to the difference in techniques and the large associated error bars. We also note only one of the recorded longitudinal field measurements was detected at $3\sigma$ confidence.

Using the ephemeris $\textnormal{HJD}=2454340.754 + 3.97209(4) \cdot E$ \citep{spreckley_2008} we find $\phi_{\textnormal{puls}}\!=\!0.00$ at the midpoint of observation. We adopt corresponding line mask parameters of $T_{\textnormal{eff}}\!=\!6000$\,K, $\log g\!=\!2.0$\,cm\,s$^{-2}$ and $V_{t}\!=\!4$\kms \citep{usenko_2005}. The spectacular Stokes $V$ signature of Polaris is detected at high S/N and indicates a complex magnetic topology (Fig. \ref{fig:polaris_lsd}). It looks qualitatively similar to the complex signatures seen in other late-type supergiants (e.g. $\alpha$~Per; \citealt{grunhut_2010}). Since the magnitude difference between Polaris Aa and Ab in the $V$ band is 7.2 mag \citep{evans_2008} we attribute the Stokes $V$ signal entirely to the Cepheid.

The longitudinal magnetic field measured from the LSD profiles of Polaris, $\langle B_{z}\rangle=0.59\pm 0.16$~G, corresponds to a detection at greater than $3\sigma$ confidence.  

\subsection{$\eta$ Aql}
\begin{figure*}
	\includegraphics[scale=0.37]{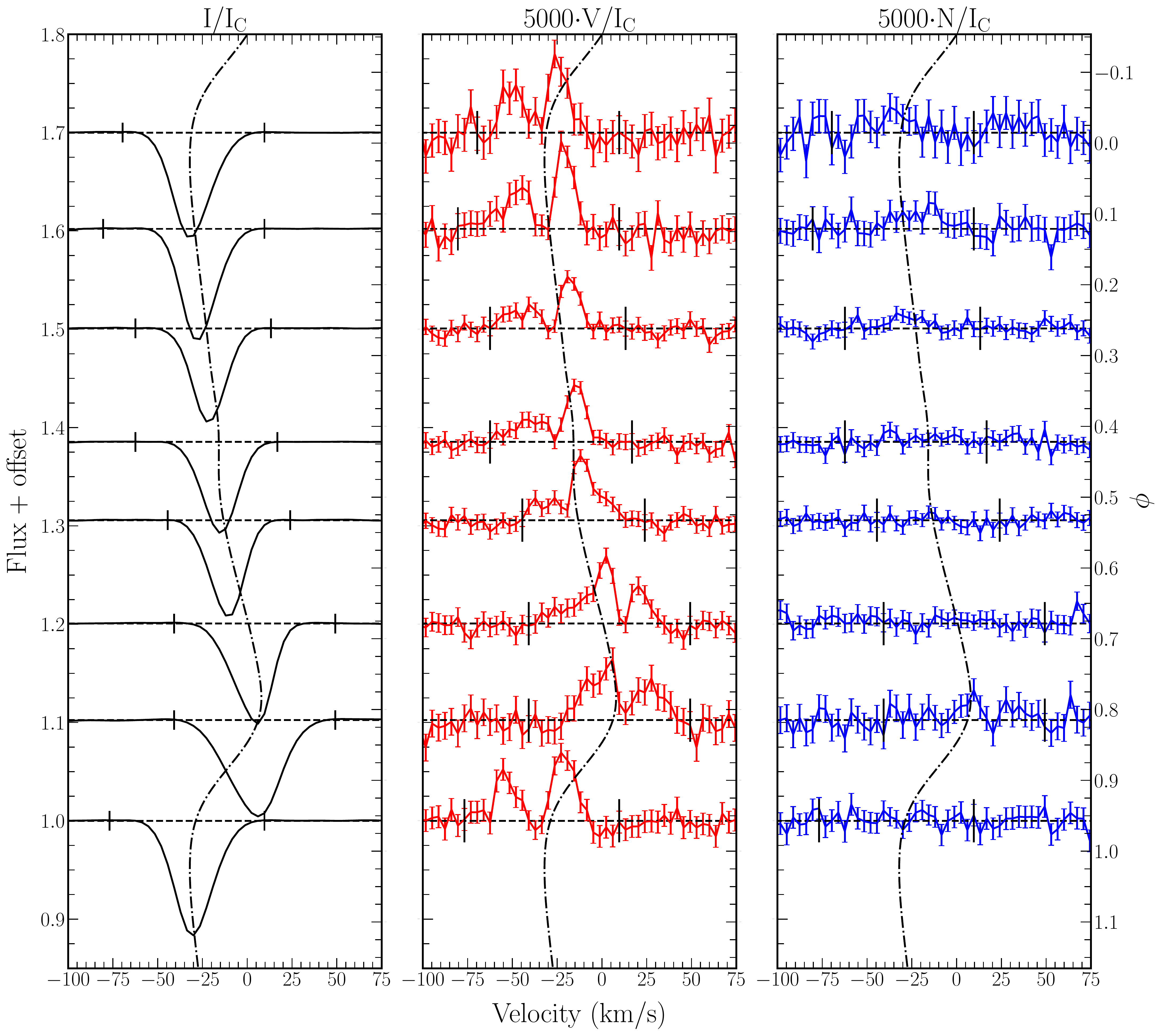}
    \caption{LSD profiles of $\eta$~Aql phased according to the ephemeris given in the text. Left Stokes~$I$ profiles (black), middle Stokes~$V$ (red) and right $N$ profiles (blue). Solid vertical lines show velocity bounds used in the magnetic analysis. Observations were obtained on consecutive nights from top to bottom. Black dash-dotted lines show $\eta$~Aql's RV curve \citep{evans_2015RV}.}
    \label{fig:etaAql_LSD}
\end{figure*}
$\eta$ Aql ($P_{\textnormal{puls}}\!=\!7.18$\,d, $\langle V\rangle\!=\!3.90$, F6Ib-G4Ib) has been well studied due to its brightness and proximity. It has two known companions, a F1V-F6V star at 0.65" detected in VLT/NACO imaging \citep{gallene_2014} and a closer unresolved B9.8V companion detected in IUE spectra \citep{evans_1991}. No X-ray emission was detected in a single XMM-Newton observation at a phase close to maximum radius \citep{evans_2021}.

Early magnetic measurements of $\eta$~Aql were performed by \cite{borra_1981} and \cite{borra_1984}. No longitudinal magnetic field was DD with $\sim\!10$\,G uncertainties. \cite{plachinda_2000} reported the detection of a longitudinal magnetic field in $\eta$~Aql, and claimed that the field varies approximately sinusoidally with pulsation phase ($-100\!<\!\langle B_{z}\rangle\!<\!+50$\,G). A discontinuity in the field variation was found at about phase 0.6 and was attributed to the propagation of a shock wave through the atmosphere. Subsequent observations were obtained by \cite{wade_2002} to test the reproducibility of these findings using the MuSiCoS spectropolarimeter at Pic du Midi observatory. The authors found no evidence of a magnetic field in the Stokes $V$ spectra and concluded that $\eta$~Aql is non-magnetic at the level of $\sim\!10$\,G. A signature in the LSD Stokes $V$ profile of $\eta$~Aql was later detected by \cite{grunhut_2010}. The Stokes $V$ profile has an unusual shape, consisting of a single positive circular polarization lobe. \cite{grunhut_2010} measured $\langle B_{z}\rangle=-0.23\pm0.75$ at $\phi_{\textnormal{puls}}=0.76$, consistent with the $<10$\,G limit placed by \cite{wade_2002}.

We calculate the phase of each observation according to the ephemeris $\textnormal{JD}=2448069.8905+7.176841(12)\cdot E$ \citep{merand_2015}. We adopt mean stellar parameters for the initial line mask of $T_{\textnormal{eff}}=5750$\,K, $\log g=2.0$\,cm s$^{-2}$ and $V_{t}=3$\kms\  \citep{luck_2018}. Figure \ref{fig:etaAql_LSD} shows the LSD profiles of $\eta$~Aql plotted as a function of pulsation phase. The Stokes $I$ profiles vary during the pulsation, and their asymmetry is due to a combination of the pulsation velocity, rotation, turbulence and limb darkening \citep{nardetto_2006}. The Stokes $V$ profiles are DD at all phases and consist predominantly of two positive lobes whose line centre moves with the radial velocity. The blue wing of several Stokes $V$ profiles show excess flux in the Stokes $I$ continuum. The shape of the Stokes $V$ profiles do not change dramatically across observations leading us to conclude that the magnetic field remains relatively stable during pulsation. The first observation at $\phi_{\textnormal{puls}}\!=\!0.99$ and the last observation at $\phi_{\textnormal{puls}}\!=\!0.96$ have similar Stokes $V$ signatures. This suggests that most of the observed Stokes $V$ variation is driven by pulsation, not rotation.

Additionally, we identified three sets of archival Stokes~$V$ spectra of $\eta$~Aql including the magnetic detection by \cite{grunhut_2010}\footnote{LSD profiles of the archival observations are shown in Appendix~\ref{sect_archival}.}. To the best of our knowledge the other two archival spectra are unpublished. We re-analyzed the 2009 ESPaDOnS Stokes~$V$ spectra obtained by \cite{grunhut_2010} using our prepared line mask. The LSD Stokes~$V$ profile is DD with a comparable morphology and $\langle B_{z}\rangle$ measurement to that found by \cite{grunhut_2010}.

The second archival observation was obtained in 2011 with ESPaDOnS (ID: 11AC18) and has a DD null signature that is mirrored in Stokes $V$. This significant null feature indicates that the spectrum is affected by systematics. As we are unable to extract anything physically meaningful from the LSD Stokes~$V$ profile we omit this observation from Table~\ref{tab:obs} and Appendix~\ref{sect_archival}. 
The third archival observation of $\eta$~Aql was acquired in 2015 with NARVAL (ID: L151N09). The Stokes $V$ profile is ND with a longitudinal field uncertainty of 0.6\,G. The ND is consistent with our newer observation at similar phase ($\phi_{\textnormal{puls}}=0.81$) given the lower S/N (710 vs. 3370 at 500 nm).

\subsection{$\zeta$ Gem}
\begin{figure*}
	\includegraphics[width=\textwidth]{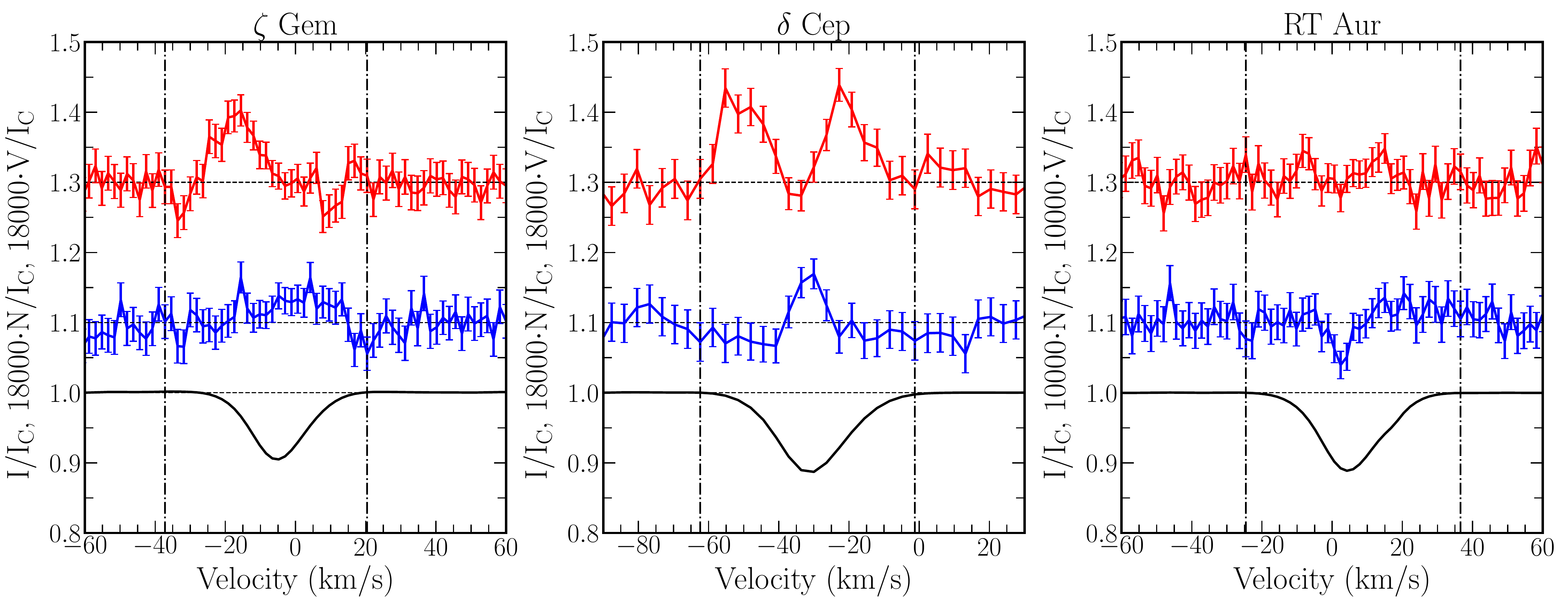}
    \caption{Same as Fig. \ref{fig:polaris_lsd} for $\zeta$ Gem, $\delta$ Cep and RT Aur. Note the difference in scaling of the LSD Stokes~$V$ and $N$ profiles compared to Polaris.}
    \label{fig:3_lsdplot}
\end{figure*}
$\zeta$ Gem ($P_{\textnormal{puls}}=10.15$\,d, $\langle V\rangle=3.92$, F7Ib-G3Ib) has the longest pulsation period of our targets observed to date. No clear evidence of a close binary companion has been detected in RV variations \citep{evans_2015RV}, $\textit{Gaia}$ observations \citep{kervella_2019} or long-baseline interferometry \citep{Breitfelder_2016}. X-ray emission has been tentatively detected at a phase close to maximum radius \citep{bohm_1983}, consistent with observations of $\delta$~Cep \citep{engle_2017}. No magnetic field has been clearly detected in $\zeta$~Gem to date \citep{borra_1981, borra_1984, grunhut_2010}.

Using the ephemeris $\textnormal{JD}=2449135.061+10.149806(17)\cdot E$ \citep{Breitfelder_2016} we calculate $\phi_{\textnormal{puls}}=0.07$ and adopt corresponding line mask parameters of $T_{\textnormal{eff}}=5750$\,K, $\log g=1.5$\,cm\,s$^{-2}$ and $V_{t}=3$\kms \citep{luck_2018}. The Stokes~$V$ profile (Fig. \ref{fig:3_lsdplot}) is DD and we measure $\langle B_{z}\rangle = 0.41\pm0.16$\,G. The Stokes $V$ profile consists of a positive circular polarization lobe similar to that seen previously in $\eta$~Aql \citep{grunhut_2010}. There is excess Stokes~$V$ flux in the blue wing of the profile that extends into the Stokes~$I$ continuum.

We re-analyzed the 2009 archival ESPaDOnS spectra of \cite{grunhut_2010} using our prepared line mask and similarly found no detection with a comparable $\langle B_{z}\rangle$ measurement. The increased S/N (1750 vs. 4700 at 500\,nm) due to the longer integration time ($\sim\!6.5$ times longer) has permitted the detection of a magnetic signature (see Appendix~\ref{sect_archival}).

\subsection{$\delta$ Cep}

$\delta$ Cep ($P_{\textnormal{puls}}=5.37$\,d, $\langle V\rangle=3.95$, F5Ib-G1Ib) is the prototypical Cepheid and one of the most studied variable stars. A close orbiting companion has been detected from high precision RV measurements and is at least 100 times fainter in the optical than the Cepheid primary \citep{anderson_2015}. No magnetic field has been definitely detected in previous studies \citep{borra_1981, borra_1984, grunhut_2010}.

Using the ephemeris $\textnormal{JD}=2448305.2362421+5.3662906(61)\cdot E$ \citep{merand_2015} we calculate $\phi_{\textnormal{puls}}\!=\!0.94$ and adopt corresponding line mask parameters of $T_{\textnormal{eff}}=6500$\,K, $\log\,g=2.0$\,cm\,s$^{-2}$ and $V_{t}=3$\kms \citep{luck_2018}. The Stokes $V$ profile (Fig. \ref{fig:3_lsdplot}) is DD using our adopted velocity binning of 3.6\kms and MD at the default binning of 1.8\kms. The Stokes $V$ profile consists of two distinct positive lobes qualitatively similar to those of $\eta$~Aql (Fig. \ref{fig:etaAql_LSD}). The inferred longitudinal field is $\langle B_{z}\rangle = 0.43\pm0.19$\,G.

We re-analyzed the 2009 ESPaDOnS Stokes $V$ spectra of $\delta$~Cep from \cite{grunhut_2010} and similarly found no detection. As in the case of $\zeta$~Gem, the increased S/N (1280 vs. 6530 at 500\,nm) allows us to detect a magnetic signature. In addition, we analyzed two sets of unpublished archival Stokes $V$ spectra for $\delta$~Cep obtained in 2011 with ESPaDOnS (ID: 11AC18) and in 2016 with NARVAL (ID: L161N97). We detect no magnetic signature in either observation with longitudinal field uncertainties of $\sim1$\,G. Given their lower S/N, the data are consistent with our newer observations (see Appendix~\ref{sect_archival}). 

\subsection{RT Aur}
RT Aur (48 Aur, $\langle V\rangle\!=\!5.45$, F4Ib-G1Ib, $P_{\textnormal{puls}}\!=\!3.73$\,d) is a fundamental mode pulsator, with a precisely repeating light curve in \textit{MOST} photometry \citep{evans_2015}. Its binary status is unclear as no orbital velocity variations have been detected \citep{evans_2015}, however a possible companion has been reported from long-baseline interferometry \citep{gallenne_2015}. Using the ephemeris $\textnormal{JD}=2448028.178+3.728305(5)\cdot E$ \citep{Breitfelder_2016} we calculate $\phi_{\textnormal{puls}}=0.09$ and adopt line mask parameters of $T_{\textnormal{eff}}\!=\!6500$\,K, $\log g\!=\!2.0$\,cm\,s$^{-2}$ and $V_{t}\!=\!3$\kms \citep{luck_2018}. No signal is detected in the LSD Stokes~$V$ profile (Fig.~\ref{fig:3_lsdplot}), and $\langle B_{z}\rangle = 0.03\pm0.24$\,G is consistent with zero. Increasing the velocity binning above 1.8\kms\ does not yield a detection.

\section{Origin of the Stokes~$V$ features}
In the context of a magnitude-limited survey, we report new high-S/N ESPaDOnS observations of 5 classical Cepheids: $\alpha$~UMa (Polaris), $\delta$~Cep, $\zeta$~Gem, $\eta$~Aql, and RT~Aur. Of these, a significant Stokes~$V$ signal is detected in the LSD profiles of all but one. In addition, we discuss archival observations of several of these stars. No signal is detected in the LSD profiles of the lower S/N archival observations.

Apart from Polaris, the detected Stokes~$V$ features have unusual morphologies: typically, they are approximately unipolar, exhibiting positive net circular polarization. Furthermore, those of $\eta$~Aql and $\delta$~Cep appear double-lobed. The amplitudes relative to the Stokes $I$ core depth are $V/I\sim 5\times 10^{-5}$. These amplitudes and morphologies are comparable to those of some Am stars, in which weak Stokes $V$ features have been detected using ESPaDOnS and its twin NARVAL (Sirius A; \citealt{petit_2011}; $\beta$~UMa and $\theta$~Leo; \citealt[][]{blazere_2016}). \citet{2014psce.conf..389K} reported the confirmation of these signatures using a spectropolarimeter of different design (specifically HARPSpol observing Sirius~A). This argues against a persistent instrumental artifact and supports the interpretation as astrophysical in origin. %

As pointed out by \citet{blazere_2016}, a number of cool active stars have been reported to display weak net circular polarization, sometimes accompanied by centre-of-gravity shifts, in their LSD Stokes~$V$ profiles \citep[e.g.][]{2005MNRAS.361..837P,2014A&A...561A..85L,2015MNRAS.446.1988S,mathias_2018}. These features have been successfully interpreted using concepts from solar physics, where analogous Stokes~$V$ features are routinely described \citep[e.g.][]{1993SSRv...63....1S} and are explained by vertical gradients in the velocity field and magnetic field strength \citep[][and references therein]{2002ApJ...564..379L}. \citet{blazere_2016} proposed that a similar explanation may apply to the Am stars exhibiting Stokes $V$ signatures. Indeed, preliminary modelling using 1D polarized radiative transfer calculations \citep{folsom_2018} successfully reproduced many characteristics of the Stokes~$V$ profile of Sirius~A. Cepheids are inherently pulsating stars, and their dynamic atmospheres are well known to exhibit strong velocity gradients. Coupling this with the fact of their cool, convective atmospheres that are conducive to dynamo activity, it seems natural in retrospect to expect Stokes~$V$ profiles with the properties we have observed. 

On the other hand, the line profile variations associated with pulsation are known to potentially complicate magnetic analysis of pulsating stars \citep[e.g.][]{neiner_2012,2018MNRAS.478L..39S} and even introduce spurious detections \citep[e.g.][]{wade_2002,2004A&A...413.1087C}. It is typically advised that spectropolarimetric sequences span no more the 1/20th of a star's pulsation cycle (i.e. 5\% in pulsation phase) to avoid these effects. In our sample, the pulsation periods of the detected targets range from about 4\,d to about 10\,d, and the corresponding relative duration of the spectropolarimetric sequences was at most 2.5\% in pulsation phase. Moreover, we have detected similar signatures in stars that span a large range of pulsation period. In particular we obtain no detection in RT~Aur - notwithstanding that this star has the shortest pulsation period and longest exposure time of our sample, and that it was observed with a similar magnetic precision as the others. Therefore, it seems unlikely that instrumental systematics due to pulsation cause the peculiar signatures.

In the cases of some cool evolved stars, the presence of strong line linear polarization in the presence of weak instrumental crosstalk was investigated as a potential origin of/contributor to the Stokes~$V$ features (e.g. $\chi$~Cyg; \citealt{2014A&A...561A..85L}; Betelgeuse; \citealt{mathias_2018}). To test this scenario we obtained Stokes~$Q$ and $U$ spectra for all of the detected targets in this paper during a recent ESPaDOnS run. No linear polarization was detected. These results are summarized in Appendix~\ref{sect_linear}.

We tentatively conclude that the Stokes~$V$ features that we detect in LSD profiles of classical Cepheids are likely not a consequence of instrumental or observational systematics, and that a dynamo-generated magnetic field coupled with atmospheric velocity and magnetic gradients provides a reasonable and natural qualitative explanation of their properties. 

\section{Summary and next steps}
This survey was designed to provide a first look into the incidence and properties of Cepheid magnetic fields, with longer-term goals to study the relationship of Cepheid magnetism to atmospheric structure and dynamics and subsequent evolution. 

We have clearly detected Stokes~$V$ features in 4 of 5 Cepheids observed. As described above, the features exhibit a diversity of morphologies. That of Polaris is most easily understood in the context of the normal Zeeman effect. Those of the other stars are peculiar in that they are roughly unipolar, and appear to extend somewhat outside the visible bounds of their associated LSD Stokes~$I$ profile. Based on their similarity in morphology and amplitude to Stokes~$V$ features detected in LSD profiles of other cool stars and some Am stars, we tentatively attribute them to the Zeeman effect influenced by vertical velocity and magnetic field gradients. For one of these stars ($\eta$~Aql) we have repeatedly detected Stokes~$V$ features across a complete pulsation cycle. Considering the first detection of Stokes~$V$ features in LSD profiles of this star by \citet{grunhut_2010}, this demonstrates that the detectability of these features is a long-lived, repeatable phenomenon. 

The Stokes $V$ profile of Polaris is remarkable. It is complex, detected at high S/N, and distinct from those of the other detected targets. Given the importance of Polaris as the brightest and most nearby Cepheid, this is a result of great significance. Immediate magnetic monitoring of this key star has been initiated, with the objective to identify its rotational period and map its magnetic field using Zeeman-Doppler Imaging.

Given the dynamic nature of Cepheid atmospheres and the clear presence of velocity gradients and possibly magnetic gradients, the modeling initiated by \citet{folsom_2018} seems likely to provide a fruitful context to explore the origin of the peculiar profiles of $\zeta$~Gem, $\delta$~Cep, and $\eta$~Aql. This will also be a key next step in our investigation that may additionally provide a new probe of Cepheid atmospheric properties.

The observations presented in this paper quadruple the number of known magnetic Cepheids and suggest that magnetic fields, likely of dynamo origin, are commonly detectable in this class of star when observed with sufficient precision. 

\section*{Acknowledgements}
We thank the referee, Dr. Stefano Bagnulo, for thoughtful comments leading to the improvement of this paper. GAW acknowledges support in the form of a Discovery Grant from the Natural Science and Engineering Research Council (NSERC) of Canada. This research has made use of the SIMBAD database, operated at CDS, Strasbourg, France. This work has made use of the VALD database, operated at Uppsala University, the Institute of Astronomy RAS in Moscow, and the University of Vienna.

\section*{Data Availability}
The data underlying this paper are available from the Canadian Astronomy Data Centre (https://www.cadc-ccda.hia-iha.nrc-cnrc.gc.ca) and PolarBase (http://polarbase.irap.omp.eu) after the conclusion of the proprietary period. The data can be accessed with the target names given in Table \ref{tab:obs} and program IDs given in text.

\bibliographystyle{mnras}
\bibliography{example} %

\newpage
\clearpage
\appendix
\section{Archival Observations}\label{sect_archival}
\begin{figure*}[t]
	\includegraphics[width=\textwidth]{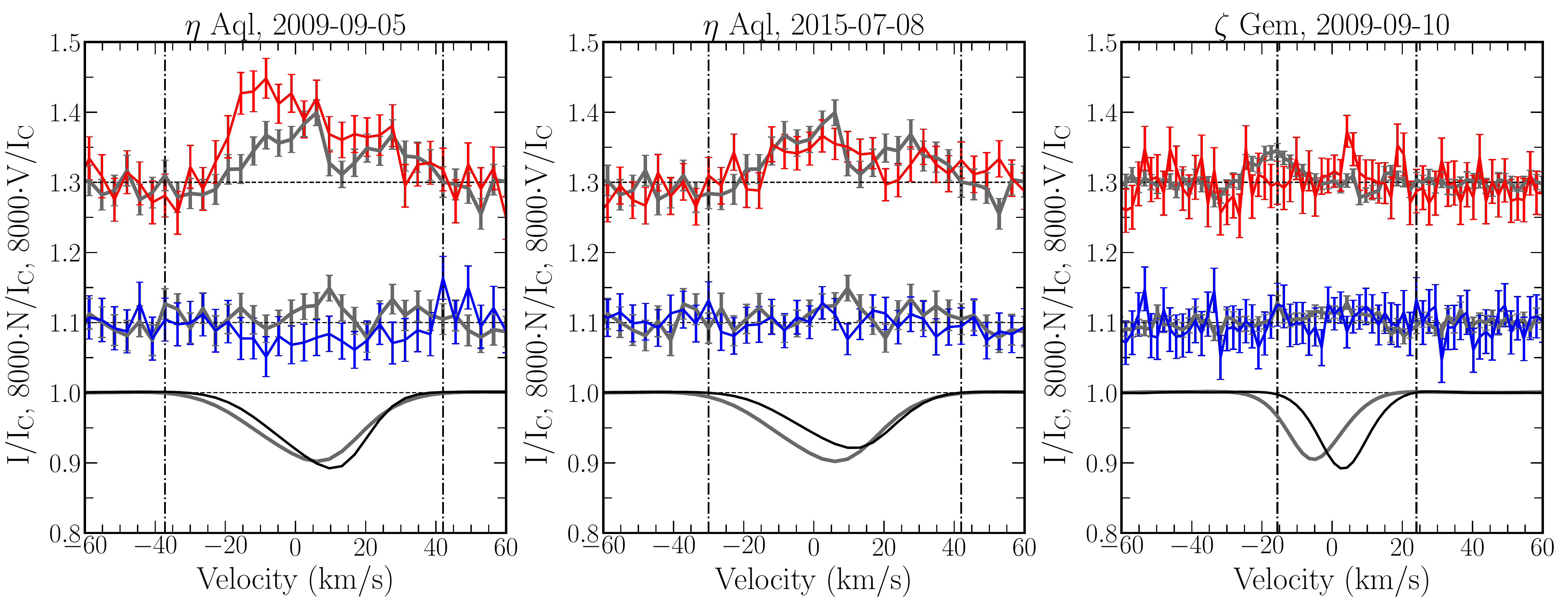}
    \caption{LSD Stokes~$V$ (top, red), diagnostic null $N$ (middle, blue) and Stokes~$I$ (bottom, black) profiles from archival observations of $\eta$~Aql and $\zeta$~Gem. The thick grey lines denote the LSD profiles of the magnetic detections shown in Fig. \ref{fig:etaAql_LSD} ($\eta$~Aql, $\phi_{\textnormal{puls}}=0.81$) and Fig. \ref{fig:3_lsdplot} ($\zeta$~Gem). Vertical dash-dotted lines show the velocity bounds used in the magnetic analysis. The Stokes~$V$ profile of the 2009 observation of $\eta$~Aql is DD (left). The Stokes~$V$ profile consists of a single positive lobe that is consistent with the findings of \protect\cite{grunhut_2010}. The Stokes~$V$ profile of the 2015 NARVAL observation of $\eta$~Aql is formally ND (middle), although we note by eye there does appear to be a marginal positive circular polarization signal. The pixel uncertainties of the NARVAL Stokes~$V$ profile are approximately 20\% larger than our detected signature at similar phase. The Stokes~$V$ profile of the 2009 ESPaDOnS observation of $\zeta$~Gem is ND (right), consistent with the results of \protect\cite{grunhut_2010}. The amplitude of the detected Stokes~$V$ profile of $\zeta$~Gem (grey) is comparable to the noise level of the archival profile with $\sim3$ times smaller pixel uncertainties. This illustrates that the Stokes~$V$ features detected in our recent higher S/N observations would not have been detected in the lower S/N archival data.}
    \label{fig:etaAql_zetaGem_archival}
\end{figure*}
\begin{figure*}[t]
	\includegraphics[width=\textwidth]{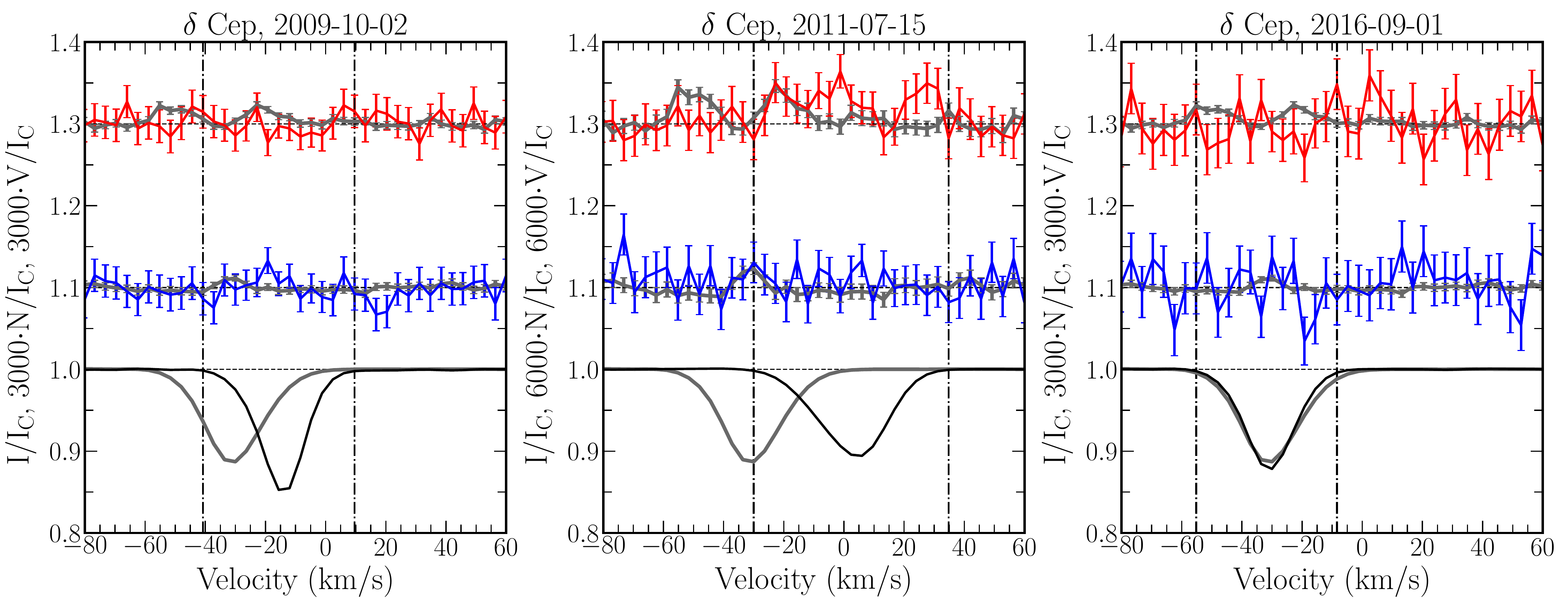}
    \caption{Same as Fig. \ref{fig:etaAql_zetaGem_archival} for archival observations of $\delta$~Cep. The thick grey lines denote the LSD profiles of the magnetic detection of $\delta$~Cep shown in Fig.~\ref{fig:3_lsdplot}. Note that the 2011 LSD Stokes $V$ and $N$ profiles have a higher scaling for visibility. The amplitude of the detected Stokes~$V$ profile of $\delta$~Cep is comparable to or smaller than the noise level of the archival profiles with $\sim$3-7 times smaller pixel uncertainties. These results illustrate that the Stokes~$V$ features detected in our recent higher S/N observations would not have been detected in the lower S/N archival data.}
    \label{fig:delCep_archival}
\end{figure*}
\clearpage
\newpage
\section{LSD Stokes $Q$ and $U$ Profiles}\label{sect_linear}
For weak magnetic fields ($<$1 kG), linear polarization amplitudes due to the Zeeman effect are expected to be about an order of magnitude smaller than circular polarization amplitudes. However, the LSD Stokes $Q$ and $U$ amplitudes of some evolved stars are found to be about an order of magnitude higher than the LSD Stokes $V$ amplitudes. These stars include $\chi$~Cyg, \citep{2014A&A...561A..85L}, Betelgeuse \citep{auriere_2016, mathias_2018} and $\mu$~Cep \citep{tessore_2017}. In the case of Betelgeuse, this comparatively strong linear polarization has been attributed to convective cell brightness spots in the photosphere of the star and Rayleigh scattering in the continuum \citep{lopez_2018}.     

Since ESPaDOnS is affected by weak ($<1\%$; \citealt{2010SPIE.7735E..4CB}) polarimetric crosstalk, the presence of strong linear polarisation in the spectra of our targets could produce spurious Stokes $V$ features. To test this possibility we observed a number of our targets in linear polarization with ESPaDOnS. 

In a typical magnetic analysis, the LSD Stokes $Q$ and $U$ profiles are generated using a different weighting than LSD Stokes $V$ profiles \citep{2000MNRAS.313..823W,1979A&A....74....1R}. However, as we are investigating possible contamination of LSD Stokes $V$ profiles by linear polarization of a non-Zeeman origin we apply the same LSD procedure discussed in Sect. \ref{sec:magnetic_diagnosis} to the Stokes $Q$ and $U$ spectra\footnote{Since we seek to test if the detected LSD Stokes~$V$ profiles are a consequence of cross-talk from strong linear polarization signatures, it is appropriate to apply the Stokes~$V$ LSD scaling in our analysis. Nevertheless, for completeness we also applied the LSD procedure to Stokes~$Q$ and $U$ spectra using four additional weighting schemes, scaling by: (1) line depth $d$, (2) wavelength $\lambda$, (3) line depth and wavelength $d\lambda$, (4) line depth, wavelength squared and Landé factor squared $d\lambda z^{2}$ (scaling appropriate to Zeeman Stokes Q and U profiles; \citealt{wade_2000}). We did not detect any signal above noise in any of these tests}.

The log of linear polarization observations is reported in Table~\ref{tab:obs_linear}. The S/N of the Stokes~$Q$ and $U$ spectra at 500\,nm is $\sim5$ times lower than the corresponding Stokes~$V$ spectra (see Table~\ref{tab:obs}). This lower S/N is sufficient to allow us to rule out cross-talk contamination as ESPaDOnS cross-talk was last measured to be $0.1\%$ in November 2017\footnote{\label{cross-talk}https://www.cfht.hawaii.edu/Instruments/Spectroscopy/Espadons/}. If the Stokes~$V$ signatures we detect were due to cross-talk, we would expect to see Stokes~$Q$ and $U$ profiles with amplitudes of order $100-1000$ times larger than the Stokes~$V$ profiles. Linear polarization signals of this amplitude would be clearly detected at the S/N of our Stokes~$Q$ and $U$ spectra.

\begin{figure*}
	\includegraphics[width=\textwidth]{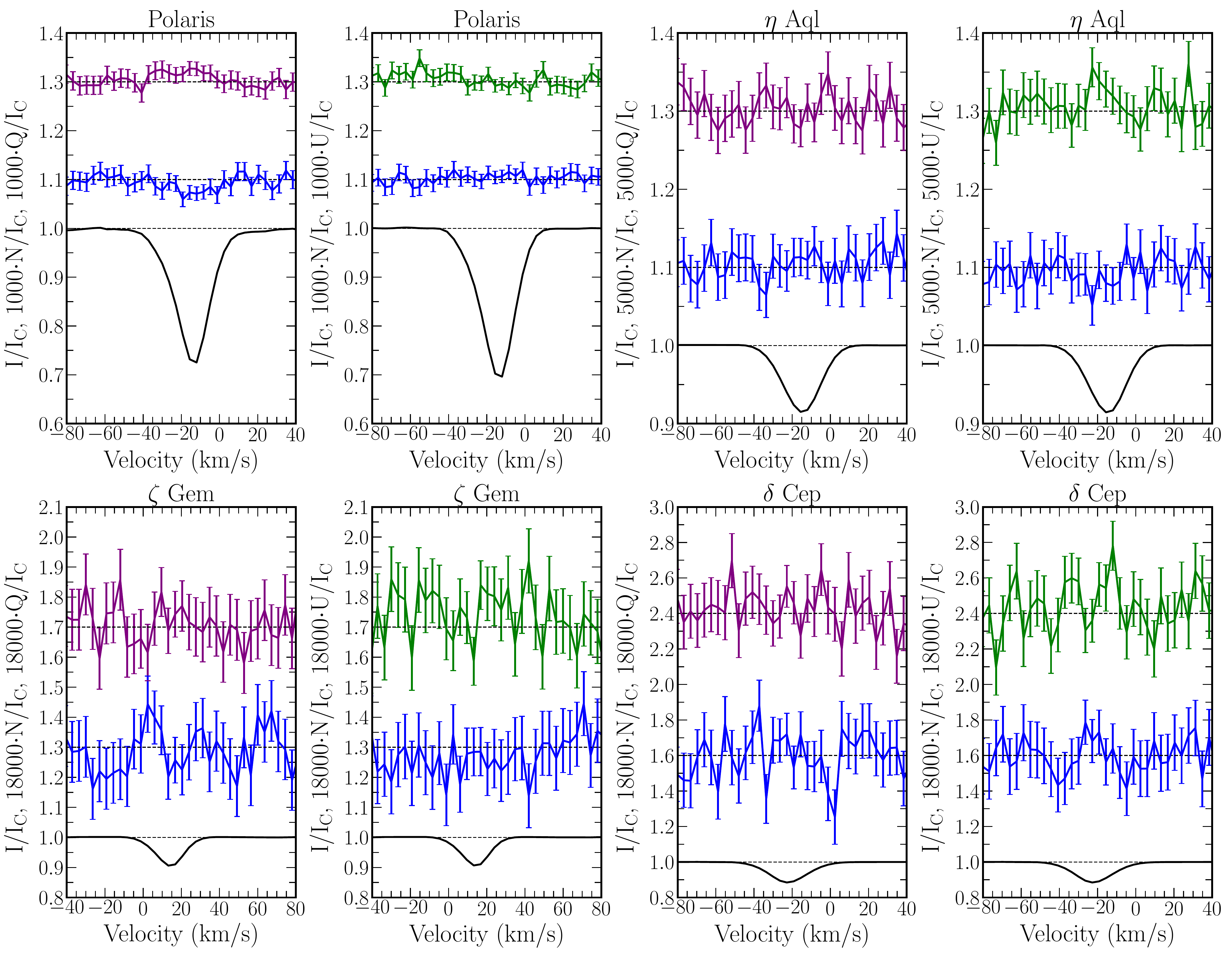}
    \caption{LSD Stokes~$Q$ and $U$ (top, green / purple), diagnostic null $N$ (middle, blue) and Stokes~$I$ (bottom, black) profiles of the four Cepheids with magnetic detections. The LSD $Q$ and $U$ profiles are generated using the same line masks and procedure as the LSD Stokes~$V$ profiles in Figs. \ref{fig:etaAql_LSD} and \ref{fig:3_lsdplot} with a velocity binning of 3.6\kms. No net linear polarization is detected.}
    \label{fig:linear}
\end{figure*}

As illustrated in Fig.~\ref{fig:linear}, no linear polarization was detected in the LSD Stokes~$Q$ and $U$ profiles. Therefore, we rule out line linear polarization of sufficient strength to explain the observed Stokes~$V$ features.

\begin{table}
\caption{Log of ESPaDOnS linear spectropolarimetric observations obtained on 2021-11-23. Included are each target's ID, Stokes parameter, heliocentric Julian date at midpoint of observation, pulsation phase $\phi_{\textnormal{puls}}$ at midpoint of observation (adopted ephemerides described in text), exposure time of polarimetric sequence and S/N per 1.8\kms\ pixel at 500\,nm of the co-added spectrum.}
\label{tab:obs_linear}
\begin{tabular}{cccccc}
\hline
Star&Stokes&HJD\,$-$&$\phi_{\textnormal{puls}}$&Exp. Time&S/N\\
& &2,400,000&&(s)&\\
\hline
Polaris&Q&59541.82642&0.40&$2\times4\times11$&1110\\
&U&59541.83019&0.41&$2\times4\times11$&1160\\
$\eta$ Aql&Q&59541.68223&0.45&$1\times4\times50$&900\\
&U&59541.68619&0.45&$1\times4\times50$&910\\
$\zeta$ Gem&Q&59541.88057&0.32&$1\times4\times50$&790\\
&U&59541.88458&0.32&$1\times4\times50$&810\\
$\delta$ Cep&Q&59541.70662&0.90&$1\times4\times50$&1110\\
&U&59541.71051&0.90&$1\times4\times50$&1030\\
\hline
\end{tabular}
\end{table}

\bsp	%
\label{lastpage}
\end{document}